\documentclass[11pt]{article}  
\usepackage{latexsym}
\usepackage{graphicx}
\newcommand{\myvev}[1]{{\langle #1 \rangle}}
\newcommand{\vev}{{\it vev}}

\def\rvev{\right\rangle}
\def\lvev{\left\langle}
\def\thefootnote{\fnsymbol{footnote}}
\def\Ph{\bar{\Phi}}
\def\Tr{{\rm Tr}}
\def\im{{\rm Im}}
\def\re{{\rm Re}}
\def\dx{\delta_X}
\def\ux{$U(1)_X$}
\def\ua{$U(1)_a$}

\def\t{\bar{t}}
\def\[{\left [}
\def\]{\right ]}
\def\({\left (}
\def\){\right )}
\def\bl{\bar{\lambda}}

\def\pp{\partial}

\def\G{{\cal G}}

\newcommand{\pl}{{\it Phys.\ Lett. }}
\newcommand{\np}{{\it Nucl.\ Phys. }}

\def\L{{\cal L}}

\def\Tev{{\rm TeV}}
\def\Gev{{\rm GeV}}
\newcommand{\beq}{\begin{equation}}
\newcommand{\eeq}{\end{equation}}
\newcommand{\bear}{\begin{eqnarray}}
\newcommand{\eear}{\end{eqnarray}}

\begin{document}
\begin{titlepage}

      \hfill  LBNL-53299

      \hfill  UCB-PTH-03/16

      \hfill hep-ph/0307124

\hfill\today
\\[.2in]

\begin{center}

{\large \bf Supergravity from the weakly coupled heterotic
string}\footnote{Talk presented at Sugra20, Northeastern University,
Boston, MA, March 17--21, 2003, to be published in the
proceedings.}\footnote{This work was supported in part by the
Director, Office of Energy Research, Office of High Energy and Nuclear
Physics, Division of High Energy Physics of the U.S. Department of
Energy under Contract DE-AC03-76SF00098 and in part by the National
Science Foundation under grants PHY-0098840 and INT-9910077.}

Mary K. Gaillard \\[.05in]

{\em Department of Physics,University of California, and Theoretical
 Physics Group, 50A-5101, Lawrence Berkeley National Laboratory,
 Berkeley, CA 94720, USA}\\[.2in]
\end{center}

\begin{abstract} The weakly coupled vacuum of $E_8\otimes E_8$ heterotic
string theory remains a promising candidate for the underlying theory
of the Standard Model.  The particle spectrum and the issue of dilaton
stabilization are reviewed.  Specific models for hidden sector
condensation and supersymmetry breaking are described and their
phenomenological and cosmological implications are discussed.

\end{abstract}
\end{titlepage}

\newpage
\renewcommand{\thepage}{\roman{page}}
\setcounter{page}{2}
\mbox{ }

\vskip 1in

\begin{center}
{\bf Disclaimer}
\end{center}

\vskip .2in

\begin{scriptsize}
\begin{quotation}
This document was prepared as an account of work sponsored by the United
States Government.  While this document is believed to contain
correct information, neither the United States Government nor any agency
thereof, nor The Regents of the University of California, nor any of their
employees, makes any warranty, express or implied, or assumes any legal
liability or responsibility for the accuracy, completeness, or usefulness
of any information, apparatus, product, or process disclosed, or represents
that its use would not infringe privately owned rights.  Reference herein
to any specific commercial products process, or service by its trade name,
trademark, manufacturer, or otherwise, does not necessarily constitute or
imply its endorsement, recommendation, or favoring by the United States
Government or any agency thereof, or The Regents of the University of
California.  The views and opinions of authors expressed herein do not
necessarily state or reflect those of the United States Government or any
agency thereof, or The Regents of the University of California.
\end{quotation}
\vfill

\end{scriptsize}

\vskip 2in

\begin{center}
\begin{small}
{\it Lawrence Berkeley Laboratory is an equal opportunity employer.}
\end{small}
\end{center}

\newpage
\renewcommand{\theequation}{\arabic{section}.\arabic{equation}}
\renewcommand{\thepage}{\arabic{page}}
\setcounter{page}{1}
\def\thefootnote{\arabic{footnote}}
\setcounter{footnote}{0}

\section{Introduction}
In this talk I review\cite{refs} the motivation for the view that the
weakly coupled heterotic string (WCHS), with Calabi-Yau (or a CY-like
orbifold) compactification remains a prime candidate for the
description of observed physical phenomena.  Several years ago,
dilaton stabilization -- needed to fix the gauge coupling constant and
thought to be an intractable problem in this context -- was
shown\cite{us1} to be achievable by invoking nonperturbative
corrections\cite{shenk} to the dilaton K\"ahler potential. I will
review the properties of a class of models\cite{us} based on orbifold
compactifications of the heterotic string, with supersymmetry broken
by condensation in a hidden sector with a strongly coupled gauge
group.  New results, that incorporate an anomalous $U(1)$ in the
effective supergravity theory, will be presented.

\section{The case for the weakly coupled heterotic string}

\subsection{ Bottom up approach}
This approach starts from experimental data with the aim of
deciphering what it implies for an underlying, more fundamental
theory. One outstanding datum is the observed large hierarchy
between the $Z$ mass, characteristic of the scale of
electroweak symmetry breaking, and the reduced Planck scale $m_P$:
$$ m_Z\approx 90\Gev \ll m_P = \sqrt{8\pi\over
G_N}\approx2\times10^{18}\Gev, $$ which can be technically understood
in the context of supersymmetry (SUSY).  The conjunction of SUSY and
general relativity (GR) implies supergravity (SUGRA).  The absence of
observed SUSY partners (sparticles) requires broken SUSY in the
vacuum, and the observed particle spectrum constrains the mechanism of
SUSY-breaking in the observable sector: spontaneous SUSY-breaking is
not viable, leaving soft SUSY-breaking as the only option that
preserves the technical SUSY solution to the hierarchy problem.  This
means introducing SUSY-breaking operators of dimension three or
less--such as gauge invariant masses--into the Lagrangian for the SUSY
extension of the Standard Model (SM).  The unattractiveness of these
{\it ad hoc} soft terms suggests they arise from spontaneous SUSY
breaking in a ``hidden sector'' of the underlying theory.  Based on
the above facts, a number of standard scenarios have emerged.  These
include: 1) Gravity mediated SUSY-breaking, usually understood as
``minimal SUGRA'' (mSUGRA), which has been the focus of a number of
talks at this meeting. This scenario is typically characterized
by\newline $ m_{\rm scalars}= m_{0} \sim m_{\rm gravitino} =
m_{3\over2}> m_{\rm gauginos} = m_{1\over2}$ at the weak scale.
2) Anomaly mediated SUSY-breaking\cite{rs,hit}, in which $m_0
= m_{1\over2}=0$ classically; these models are characterized by $
m_{3\over2} >> m_0 ,\; m_{1\over2},$ and typically $m_0 >
m_{1\over2}$.  An exception is the Randall-Sundrum (RS) ``separable
potential'', constructed\cite{rs} to mimic SUSY-breaking on a brane
spatially separated from our own in a fifth dimension; in this
scenario $m_0^2 < 0$ and $m_0$ arises first at two loops. In general,
the scalar masses at one loop depend on the details of Planck-scale
physics.\cite{bgn} 3) Gauge mediated SUSY uses a hidden sector that has
renormalizable gauge interactions with the SM particles, and is
typically characterized by small $m_{1\over2}$.
\begin{figure}[b]
\begin{picture}(350,220)(0,20)
\put(0,36){D = 9}\put(0,116){D = 10}\put(0,196){D = 11}
\multiput(120,40)(160,0){2}{\circle{40}}
\multiput(80,120)(80,0){4}{\circle{40}} \put(200,200){\circle{40}}
\put(68,104){\line(5,-6){40}}\put(260,120){\vector(1,-3){20}}
\put(107,57){\vector(1,-1){1}}\put(293,57){\vector(-1,-1){1}}
\put(332,104){\line(-5,-6){40}}\put(140,120){\vector(-1,-3){20}}
\put(184,188){\vector(-1,-2){24}}\put(216,188){\vector(1,-2){24}}
\put(188,184){\line(0,-1){80}}\put(212,184){\line(0,-1){80}}
\put(188,104){\vector(-1,-1){52}}\put(212,104){\vector(1,-1){52}}
\put(200,40){\vector(-1,0){60}}\put(200,40){\vector(1,0){60}}
\put(262,120){\vector(-1,0){2}}\put(262,136){\vector(-1,0){10}}
\put(304,134){\vector(0,-1){2}}\put(336,134){\vector(0,-1){2}}
\put(262,128){\oval(16,16)[r]}\put(320,134){\oval(32,32)[t]}
\put(222,116){WCHS}\put(304,116){$O(32)$}
\put(298,134){I}\put(340,134){H}\put(194,196){M}
\put(116,36){II}\put(270,36){H/I}\put(72,116){IIB}\put(152,116){IIA}
\put(88,80){\circle{8}}\put(130,90){\circle{8}}\put(220,164){\line(1,0){16}}
\put(312,80){\circle{8}}\put(270,90){\circle{8}}\put(172,164){\circle{8}}
\put(162,78){\oval(8,4)}\put(162,78){\oval(16,8)}
\put(230,78){\oval(4,8)}\put(246,78){\oval(4,8)[r]}
\put(230,82){\line(1,0){16}}\put(230,74){\line(1,0){16}}
\put(180,20){$T\leftrightarrow 1/T$}\put(242,140){$T\leftrightarrow
1/T$} \put(300,156){$S\leftrightarrow 1/S$}
\end{picture}
\caption{M-theory according to John Schwarz.
\label{fig:john}}
\end{figure}
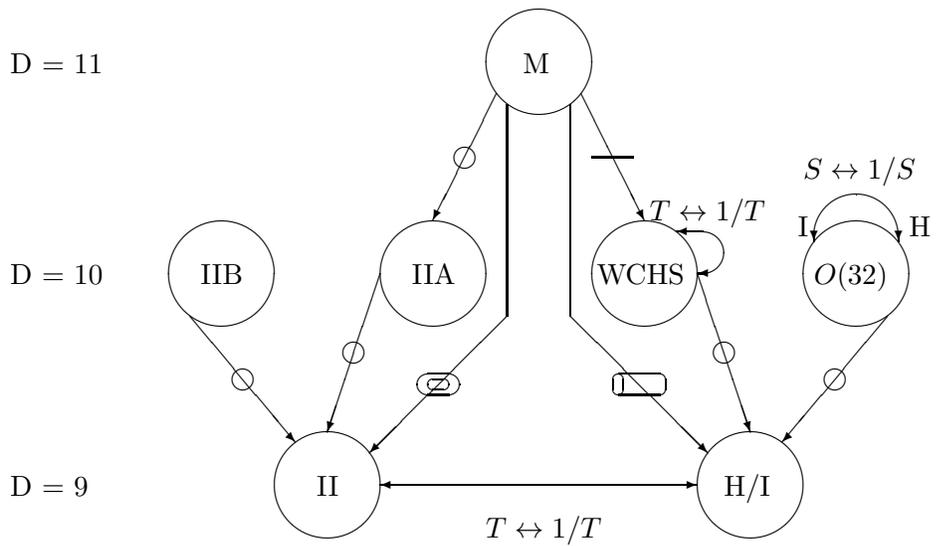
\begin{figure}[t]
\begin{picture}(350,215)(0,40)
\put(150,205){IIA,IIB: D-Branes}\put(120,200){\line(1,0){160}}
\multiput(160,190)(40,0){2}{$\times$}
\put(120,180){\oval(60,40)[l]}\put(280,190){\oval(40,20)[r]}
\put(120,140){\oval(40,40)[r]}\put(280,155){\oval(70,50)[l]}
\put(120,120){\line(-1,0){10}}\put(110,105){\oval(60,30)[l]}
\put(110,80){\oval(30,20)[r]}
\put(110,60){\oval(30,20)[l]}\put(280,110){\oval(80,40)[r]}
\put(280,90){\line(-1,0){10}}\put(270,75){\oval(40,30)[l]}
\put(270,55){\oval(40,10)[r]}\put(110,50){\line(1,0){160}}
\put(230,155){$\times$}\put(255,155){$O(32)_{\rm I}$}
\put(305,110){$\times$}\put(330,110){$O(32)_{\rm H}$}
\put(90,105){$\times$}\put(0,105){11-D SUGRA}\put(190,120){M}
\put(275,53){$\times$}\put(260,40){$E_8\otimes E_8$ WCHS}
\put(105,55){$\times$}\put(0,40){HW theory: {\small (very?) 
large extra dimension(s)}}\end{picture}
\caption{M-theory according to Mike Green.
\label{fig:mike}}\end{figure}
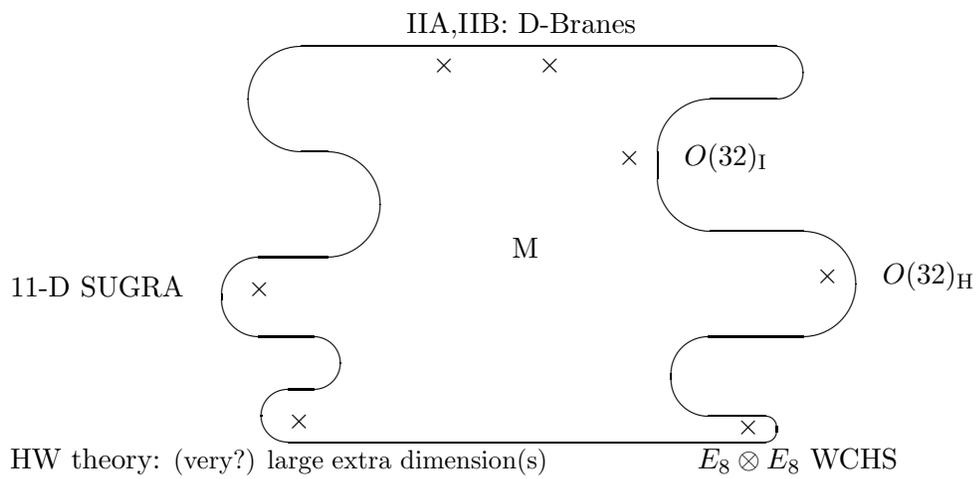

\subsection{Top down approach} 
This approach starts from a ToE with the hope of deriving the Standard
Model from it; the present prime candidate ToE is string or M theory.
The driving motivation is that superstring theory is at present the
only known candidate for reconciling GR with quantum mechanics. These
theories are consistent in ten dimensions; in recent years it was
discovered that all the consistent superstring theories are related to
one another by dualities, namely S-duality:
$\alpha\to 1/\alpha,$ and T-duality: $R\to 1/ R,$
where $\alpha$ is the fine structure constant of the gauge group(s) at
the string scale, and $R$ is a radius of compactification from
dimension D to dimension ${\rm D} -1$.  Figure~\ref{fig:john}
shows\cite{john} how these dualities relate the various 10-D
superstring theories to one another, and to the currently presumed
ToE, M-theory.  Another image of M-theory, the ``puddle diagram'' of
Figure~\ref{fig:mike}, indicates\cite{mike} that all the known
superstring theories, as well as \mbox{D $=11$} SUGRA, are particular
limits of M-theory.  Currently, there is a lot of activity in type I
and II theories, or more generally in theories with branes.  Similarly
the Ho\v rava-Witten (HW) scenario\cite{hw} and its inspirations have
received considerable attention.  If one compactifies one dimension of
the 11-D limit of M-theory, one gets the HW scenario with two 10-D
branes, each having an $E_8$ gauge group.  As the radius of this 11th
dimension is shrunk to zero, the WCHS scenario is recovered.  This is
the scenario addressed here.

\subsection{The $E_8\otimes E_8$ Heterotic String}
Here I outline the appealing aspects of the weakly coupled $E_8\otimes
E_8$ heterotic string theory.  The zero-slope (infinite string
tension) limit of superstring theory is ten dimensional supergravity
coupled to a supersymmetric Yang-Mills theory with an $E_8\otimes E_8$
gauge group.  To make contact with the real world, six of these ten
dimensions must be compact and here are assumed to be of order
$m_P\sim10^{-32}$cm.  If the topology of the extra dimensions were a
six-torus, which has a flat geometry, the 8-component spinorial
parameter of $N=1$ supergravity in ten dimensions would appear as the
four two-component parameters of $N=4$ supergravity in four
dimensions.  A Calabi-Yau (CY) manifold leaves only one of these
spinors invariant under parallel transport; the group of
transformations under parallel transport (holonomy group) is the
$SU(3)$ subgroup of the maximal $SU(4) \cong SO(6)$ holonomy group of
a six dimensional compact space.  This breaks $N=4$ supersymmetry to
$N=1$ in four dimensions.  The only phenomenologically viable
supersymmetric theory at low energies is $N=1$, because it is the only
one that admits complex representations of the gauge group that are
needed to describe quarks and leptons. For this solution, the
classical equations of motion impose the identification of the affine
connection of general coordinate transformations on the compact space
(described by three complex dimensions) with the gauge connection of
an $SU(3)$ subgroup of one of the $E_8$'s: $E_8\ni E_6\otimes SU(3)$,
resulting in $E_6\otimes E_8$ as the gauge group in four dimensions.
Since the early 1980's, $E_6$ has been considered the largest group
that is a phenomenologically viable candidate for a Grand Unified
Theory (GUT) of the SM.  Hence $E_6$ is identified as the gauge group
of the ``observable sector'', and the additional $E_8$ is attributed
to a ``hidden sector'', that interacts with the former only with
gravitational strength couplings.  Orbifolds, which are flat spaces
except for points of infinite curvature, are more easily studied than
CY manifolds, and orbifold compactifications that closely mimic CY
compactification, and that yield realistic spectra with just three
generations of quarks and leptons, have been found.\cite{iban,fiqs} In
this case the surviving gauge group is $E_6\otimes\G_o \otimes
E_8,\;\G_o\in SU(3)$.  The low energy effective field theory is
determined by the massless spectrum, {\it i.e.}, the spectrum of
states with masses very small compared with the string tension and
compactification scale. Massless bosons have zero triality under an
$SU(3)$ which is the diagonal of the $SU(3)$ holonomy group and the
(broken) $SU(3)$ subgroup of one $E_8$.  The ten-vectors $A_M,\; M =
0,1,\ldots 9,$ appear in four dimensions as four-vectors $A_\mu,\;\mu
= M = 0,1,\ldots 3$, and as scalars $A_m,\; m = M-3 = 1,\cdots 6.$
Under the decomposition $E_8\ni E_6\otimes SU(3)$, the $E_8$ adjoint
contains the adjoints of $E_6$ and $SU(3)$, and the representation
${\bf(27,3)} + {\bf(\overline{27},\overline{3})}$.  Thus the massless
spectrum includes gauge fields in the adjoint representation of
$E_6\otimes\G_o\otimes E_8$ with zero triality under both $SU(3)$'s,
and scalar fields in ${\bf 27 + \overline{27}}$ of $E_6$, with
triality $\pm1$ under both $SU(3)$'s, together with their fermionic
superpartners.  The number of ${\bf 27}$ and ${\bf\overline{27}}$
chiral supermultiplets that are massless depends on the topology of
the compact manifold.  The important point for phenomenology is the
decomposition under $E_6\to SO(10)\to SU(5)$: 
\beq\({\bf 27}\)_{E_6} =
\({\bf 16 + 10 + 1}\)_{SO(10)} = \({\bf \{\bar{5} + 10 + 1\} + \{5 +
\bar{5}\} + 1}\)_{SU(5)}.\eeq 
A ${\bf \overline{5} + 10 + 1}$ contains
one generation of quarks and leptons of the SM, a right-handed
neutrino and their scalar superpartners; a ${\bf 5 + \overline{5}}$
contains the two Higgs doublets needed in the supersymmetric extension
of the SM and their fermion superpartners, as well as color-triplet
supermultiplets. While all the states of the SM and its minimal
supersymmetric extension are present, there are no scalar particles in
the adjoint representation of the gauge group. In conventional models
for grand unification, these (or other large representations) are
needed to break the GUT group to the SM.  In string theory, this
symmetry breaking can be achieved by the Hosotani or ``Wilson line'',
mechanism in which gauge flux is trapped around ``holes'' or
``tubes'' in the compact manifold, in a manner reminiscent of the
Arahonov-Bohm effect.  The vacuum value of the trapped flux $<\int
d\ell^m A_m>$ has the same effect as an adjoint Higgs, without the
difficulties of constructing a potential for large Higgs
representations that actually reproduces the observed vacuum. When
this effect is included, the gauge group in four dimensions is \bear
&&\G_{obs}\otimes\G_{hid},
\quad\G_{obs}=\G_{SM}\otimes\G'\otimes\G_o,\quad \G_{SM}\otimes\G'\in
E_6, \quad \G_o\in SU(3),\nonumber \\ && \G_{hid}\in E_8,\quad \G_{SM}
= SU(3)_c\otimes SU(2)_L\otimes U(1)_w.
\label{eq:group}\eear

There are many other four dimensional string vacua in addition to
those described above. The attractiveness of the above picture is that
the requirement of $N=1$ SUSY naturally results in a
phenomenologically viable gauge group and particle spectrum, and the
gauge symmetry can be broken to a product group embedding the SM
without introducing large Higgs representations.  The $E_8\otimes E_8$
string theory also provides a hidden sector needed for spontaneous
SUSY-breaking.  Specifically, if some subgroup $\G_c$ of $\G_{hid}$ is
asymptotically free, with a $\beta$-function coefficient
$b_c>b_{SU(3)}$, defined by the renormalization group equation (RGE)
\beq \mu{\pp g_c(\mu)\over\pp\mu} = -{3\over2}b_cg_c^3(\mu) +
O(g_c^5)\label{eq:rge},\eeq confinement and fermion condensation will
occur at a scale $\Lambda_c\gg\Lambda_{QCD}$, and hidden sector
gaugino condensation $<\bl\lambda>_{\G_c} \ne 0,$ may
induce\cite{nilles} supersymmetry breaking.  To discuss supersymmetry
breaking in more detail, we need the low energy spectrum resulting
from the ten-dimensional gravity supermultiplet that consists of the
10-D metric $g_{MN}$, an antisymmetric tensor $b_{MN}$, the dilaton
$\phi$, the gravitino $\psi_M$ and the dilatino $\chi$.  For the class
of CY and orbifold compactifications described above, the massless
bosons in four dimensions are the 4-D metric $g_{\mu\nu}$, the
antisymmetric tensor $b_{\mu\nu}$, the dilaton $\phi$, and certain
components of the tensors $g_{mn}$ and $b_{mn}$ that form the real and
imaginary parts, respectively, of complex scalars known as moduli.
The number of moduli is related to the number of particle generations
(\# of ${\bf 27}$'s $-$ \# of ${\bf\overline{27}}$'s).  In three
generation orbifold models there are at least three moduli $t_I$ whose
$vev$'s $<{\rm Re}t_I>$ determine the radii of the three tori of the
compact space.  They form chiral multiplets with fermions $\chi^t_I$
obtained from components of $\psi_m$.  The 4-D dilatino $\chi$ forms a
chiral multiplet with with a complex scalar field $s$ whose $vev$ $
<s> = g^{-2} - i\theta/8\pi^2$ determines the gauge coupling constant
and the $\theta$ parameter of the 4-D Yang-Mills theory.  The
``universal'' axion Im$s$ is obtained by a duality transformation from
the antisymmetric tensor $b_{\mu\nu}$: $\pp_\mu{\rm
Im}s\leftrightarrow \epsilon_{\mu\nu\rho\sigma}\pp^\nu
b^{\rho\sigma}.$ Because the dilaton couples to the (observable and
hidden) Yang-Mills sector, gaugino condensation induces a
superpotential for the dilaton superfield\footnote{Throughout I use
capital Greek or Roman letters to denote a chiral superfield, and the
corresponding lower case letter to denote its scalar component.} $S$:
\beq W(S)\propto e^{-S/b_c}.\label{eq:dil}\eeq The vacuum value $
<W(S)> \propto \left<e^{-S/b_c}\right> = e^{-g^{-2}/b_c}= \Lambda_c$
is governed by the condensation scale $\Lambda_c$ as determined by the
RGE (\ref{eq:rge}).  If it is nonzero, the gravitino acquires a mass
$m_{3\over2}\propto<W>$, and local supersymmetry is broken.

\section{A model for SUSY breaking}

In this section I review the properties of a class of
models,\cite{us} based on affine level one orbifolds with three
untwisted moduli $T^I$ and a gauge group of the form (\ref{eq:group}),
with one factor $\G_c \in \G_{hid}$ that becomes strongly coupled.

\subsection{The Runaway Dilaton}

The superpotential (\ref{eq:dil}) results in a potential for the
dilaton of the form $ V(s)\propto e^{-2\re s/b_c},$ which has its
minimum at vanishing vacuum energy and vanishing gauge coupling: $<\re
s> \to\infty,\; g^2\to 0$.  This is the notorious runaway dilaton
problem.  The effective potential for $s$ is in fact determined from
anomaly matching: $\delta\L_{eff}(s,u) \longleftrightarrow
\delta\L_{hid}({\rm gauge}),$ where $u, \;\left<u\right> = \left<
\bl\lambda\right>_{\G_c},$ is the lightest scalar bound state of the
strongly interacting, confined gauge sector.  Just as in QCD, the
effective low energy theory of bound states must reflect both the
symmetries and the anomalies, {\it i.e.} the quantum induced breaking
of classical symmetries, of the underlying Yang-Mills theory.  It
turns out that the effective quantum field theory (QFT) is anomalous
under T-duality.  Since this is an exact symmetry of heterotic string
perturbation theory, it means that the effective QFT is incomplete.
This is cured by including model dependent string-loop threshold
corrections\cite{thresh} as well as a ``Green-Schwarz'' (GS)
counter-term,\cite{gsterm} analogous to the GS mechanism in 10-D
SUGRA.  This introduces dilaton-moduli mixing, and the gauge coupling
constant is now identified as
\beq g^2= 2\left<\ell\right>,\quad\ell^{-1} = 2\re s -
b{\sum_I} \ln(2\re t^I),\label{dual}\eeq
where $b\le b_{E_8} = 30/8\pi^2$ is the coefficient of the GS term,
and and $\ell$ is the scalar component of a linear superfield $L$ that
includes the two-form $b_{\mu\nu}$ and is dual to the chiral
superfield $S$ in the supersymmetric version of the
two-form/axion duality of Section 2.3.  The GS term introduces a
second runaway direction, this time at strong coupling: $V \to -
\infty$ for $g^2\to\infty$. The small coupling behavior is unaffected,
but the potential becomes negative for $\alpha = \ell/2\pi >
.57$. This is the strong coupling regime, and nonperturbative string
effects cannot be neglected; they are expected\cite{shenk} to modify
the K\"ahler potential for the dilaton, and therefore the potential
$V(\ell,u)$.  It has been shown\cite{us1} that these contributions
can indeed stabilize the dilaton.  Retaining just one or two terms of
the suggested parameterizations\cite{shenk} of the nonperturbative
string corrections: $a_n\ell^{-n/2}e^{-c_n/\sqrt{\ell}}$ or
$a_n\ell^{-n}e^{-c_n/\ell},$ the potential can be made
positive-definite everywhere and the parameters $a_n,c_n$ can be
chosen to fit two data points: the coupling constant $g^2\approx 1/2$
and the cosmological constant $\Lambda \simeq 0$.  This is fine
tuning, but it can be done with reasonable (order 1) values for the
parameters $c_n, a_n$.  If there are several condensates with
different $\beta$-functions, the potential is dominated by the
condensate with the largest $\beta$-function coefficient $b_+$, and
the result is essentially the same as in the single condensate case,
except that a small mass is generated for the axion $a=\im s$.
In these models the presence of $\beta$-function coefficients generate
mass hierarchies that have interesting implications for cosmology
and the spectrum of sparticles--the supersymmetric partners of the SM
particles.

\subsection{Sparticle Spectrum}

\begin{figure}[t]
\includegraphics{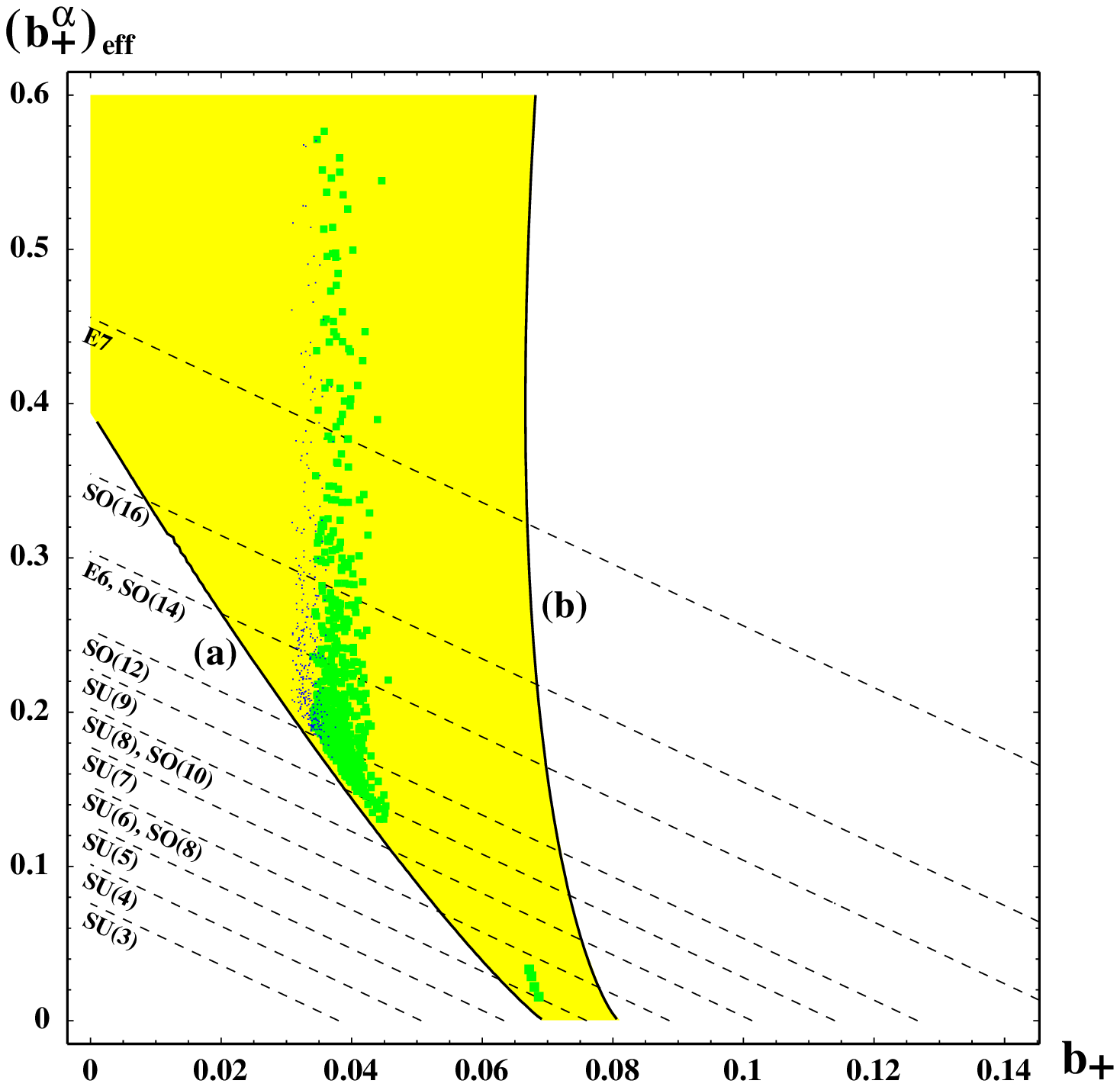}
\caption{Viable hidden sector gauge groups for scenario A of the
condensation model. The swath bounded by lines (a) and (b) is defined
by $.1<m_{3\over2}/\Tev,\lambda_c <10$, with $\lambda_c$ a
condensate superpotential coupling constant.  The fine points
correspond to $.1\le\Omega_dh^2\le.3$, and the course points to
$.3<\Omega_dh^2\le1$.  $b_c^\alpha$ is the hidden matter contribution
to $b_c.$
\label{fig:abn}}
\end{figure}

If the gauge group $\G_c\in E_8$ is smaller than $E_8$, the moduli
$t_I$ are stabilized through their couplings to twisted sector matter
and/or moduli-dependent string threshold corrections at a self-dual
point, $<t_I> = 1$ or $e^{i\pi/6}$, and their auxiliary fields vanish
in the vacuum. SUSY-breaking is dilaton mediated, avoiding a
potentially dangerous source of flavor changing neutral currents
(FCNC).  This result holds up to unknown couplings $p_A$ of chiral
matter $\phi^A$ to the GS term: at the scale $\Lambda_c$, $ m_0^A =
m_{3\over2}$ if $p_A = 0$, while $m_0^A = {1\over2}m_{t^I}\approx 10
m_{3\over2}$ if the scalar $\phi^A$ couples with the same strength as
the T-moduli: $ p_A = b$. Gaugino masses are suppressed: $m_{1\over2}
\approx b_cm_{3\over2}$ at the scale $\Lambda_c$ in the tree
approximation.  As a consequence quantum corrections can be important,
mimicking anomaly-mediated scenarios in some regions of parameter
space.  If $p_A = b$ for some gauge-charged chiral fields, there are
enhanced loop corrections to gaugino masses.\cite{gnw} Four sample
scenarios were studied\cite{gn}: A) $p_A=0$, B) $p_A=b$, C) $p_A=0$
for the superpartners of the first two generations of SM particles and
$p_A=b$ for the third, and D) $p_A=0$ for the Higgs particles and
$p_A=b$ otherwise.  Imposing constraints from experiments and the
correct electroweak symmetry-breaking vacuum rules out scenarios B and
C.  Scenario A is viable for $1.65< \tan\beta<4.5$, and scenario D is
viable for all values of $\tan\beta$, the ratio of Higgs $vev$'s in
the supersymmetric extension of the SM.  The viable range of parameter
space is shown\cite{abn} in Figure~\ref{fig:abn} for $g^2={1\over2}$.
The dashed lines represent the possible dominant condensing hidden
gauge groups $\G_c = \G_+\in E_8$ with chiral matter in the coset
space $E_8/\G_{hid}$.

\subsection{Modular Cosmology}

The masses of the dilaton $\sigma=\re s$ and the complex $t$-moduli $t^I =
\(\tau^I + i a^I\)/\sqrt{2},$ are related to the gravitino mass
by\cite{us,alex}
\beq m_\sigma \sim {1\over b_c^2} m_{3\over2}, \quad m^I_{\tau,a} \approx
{(b-b_c)\mu^I_{\tau,a}(<t^I>)\over b_c(1+b<\ell>)}m_{3\over2},
\label{eq:modmass} \eeq
where at the self-dual points $\mu^I_{\tau}\approx 3,$
$\mu^I_a\approx$ .5--1.
Taking $b = b_{E_8} \approx .38 \approx 10b_c,$ gives a
hierarchy of order $m_{3\over2}\sim 10^{-15}m_{Pl}\sim 10^3GeV$ and
$m^I_\tau\approx 30 m_{3\over2}\approx30$ TeV,
$m^I_a\approx$ (5--10)$m_{3\over2}\approx$ 5--10 TeV, $m_\sigma\sim
10^3m_{3\over2}\sim 10^6$ GeV, which is sufficient to evade the late
moduli decay problem\cite{modprob} in nucleosynthesis.

If there is just one hidden sector condensate, the axion $a = \im s$
is massless up to QCD-induced effects:
$m_a\sim(\Lambda_{QCD}/\Lambda_c)^{3\over2}m_{3\over2}\sim10^{-9}eV$,
and it is the natural candidate for the Peccei-Quinn axion. Because of
string nonperturbative corrections to its gauge kinetic term, the
decay constant $f_a$ of the canonically normalized axion is reduced
with respect to the standard result by a factor $b_c \ell^2
\sqrt{6}\approx 1/50$ if $b_c\approx .1b_{E_8}$, which may be
sufficiently small to satisfy the (looser) constraints on $f_a$ when
moduli are present.\cite{bd}

\subsection{Flat Directions in the Early Universe}

Many successful cosmological scenarios--such as an epoch of
inflation--require flat directions in the potential.  A promising
scenario for baryogenesis suggested\cite{ad} by Affleck and Dine
(AD) requires flat directions during inflation
in sparticle field space: $<\tilde q>, <\tilde\ell>\ne0$, where
$\tilde f$ denotes the superpartner of the fermion $f$.  While flat
directions are common in SUSY theories, they are generally
lifted\cite{drt} in the early universe by SUGRA couplings to the
potential that drives inflation.  This problem is evaded\cite{gmo} in
models with a ``no-scale'' structure, such as the classical potential
for the untwisted sector of orbifold compactifications. Although the
GS term breaks the no-scale property, quasi-flat
directions can still be found. An explicit model\cite{lyth} for
inflation based on the effective theory described above allows
dilaton stabilization within its domain of
attraction with one or more moduli stabilized at the vacuum value
 $t_I=e^{i\pi/6}$. One of the moduli may be the inflaton.
The moduli masses (\ref{eq:modmass}) are sufficiently large to evade the
late moduli decay problem in nucleosynthesis, but unlike the dilaton,
they are insufficient to avoid a large relic LSP density without
violation\cite{lsp} of R-parity (which distinguishes
SM particles from their superpartners). If R-parity is conserved, this
problem can be evaded if the moduli are stabilized at or near their
vacuum values--or for a modulus that is itself the inflaton.  It is
possible that the requirement that the remaining moduli be in the
domain of attraction is sufficient to avoid the problem altogether.
For example, if $\im t_I = 0$, the domain of attraction near $t_I = 1$
is rather limited: $0.6<{\rm Re}t_I<1.6$, and the entropy produced by
dilaton decay with an initial value in this range might be less than
commonly assumed.  The dilaton decay to its true ground state may
provide\cite{cgmo} partial baryon number dilution, which is generally
needed for a viable AD scenario.

\subsection{Relic Density of the Lightest SUSY Particle (LSP)}

Two pertinent questions for SUSY cosmology are: 1) Does the LSP
overclose the Universe? 2) Can the LSP be dark matter?  As discussed
by others at this meeting, the window for LSP dark matter in the
much-studied mSUGRA scenario,\cite{efo} has become smaller as the
Higgs mass limit has increased; there is not much parameter space in
which the LSP does not overclose the universe.  The ratios of
electroweak sparticle masses at the Plank scale determine the
composition of the LSP (which must be neutral) in terms of the Bino
(superpartner of the SM $U(1)$ gauge boson), the electrically neutral
Wino (superpartner of the neutral SM $SU(2)$ gauge boson), and the
higgsino (superpartner of the Higgs boson).  The mSUGRA assumption of
equal gaugino masses at the Planck scale leads to a Bino LSP with
rather weak couplings, resulting in little annihilation and the
tendency to overclose the universe, except in a narrow range of
parameter space where the LSP is nearly degenerate with the next to
lightest sparticle (in this case a stau $\tilde\tau$), allowing
significant coannihilation.  Relaxing this assumption\cite{abn}
allows a predominantly Bino LSP with a small admixture of Wino, that
can provide the observed amount $\Omega_d$ of dark matter.  In the
condensation model, this occurs in the region indicated by fine points
in Figure~\ref{fig:abn}.  Here the deviation from mSUGRA is due to
loop corrections to gaugino masses giving a small Wino component in
the LSP; its near degeneracy in mass with the lightest charged gaugino
also enhances coannihilation.  For larger $b_c$ the LSP becomes pure
Bino as in mSUGRA, and for smaller values it becomes Wino-dominated as
in anomaly-mediated models which are cosmologically safe, but do not
provide LSP dark matter, because Wino annihilation is too fast.

\section{Incorporating an anomalous $U(1)$}

Orbifold compactifications with the Wilson line/Hosotani mechanism
needed to break $E_6$ to the SM gauge group generally have $b_c\le
b\le b_{E_8}$.  An example is a model,\cite{fiqs} hereafter called
the FIQS model, with hidden gauge group $SO(10)$ and $b_c= b =
b_{SO(10)}$. It is clear from (\ref{eq:modmass}) that this would lead
to disastrous modular cosmology, since the $t$-moduli are massless.
Moreover, in many orbifold compactifications, the gauge group
$\G_{obs}\otimes\G_{hid} $ obtained at the string scale has no
asymptotically free subgroup that could condense to trigger
SUSY-breaking. However in many compactifications with realistic
particle spectra,\cite{iban,fiqs} the effective field theory
has\cite{joel2} an anomalous $U(1)$ gauge subgroup, which is not
anomalous at the string theory level.  The anomaly is
canceled\cite{dsw} by a second GS counterterm.  This results in a
D-term that forces some otherwise flat direction in scalar field space
to acquire a vacuum expectation value, further breaking the gauge
symmetry, and giving masses of order $\Lambda_D$ to some chiral
multiplets, so that the $\beta$-function of some of the surviving
gauge subgroups may be negative below the scale $\Lambda_D$, typically
an order of magnitude below the string scale.  The presence of such a
D-term was explicitly invoked in the above-mentioned inflationary
model.\cite{lyth} 

\subsection{The effective theory below the $U(1)$-breaking scale}

The GS mechanism that restores invariance under the anomalous \ux\,
gauge group induces a Fayet-Iliopoulos D-term that drives nonvanishing
\vev's for the scalar components $\phi^A$ of $n$ \ux-charged chiral
supermultiplets $\Phi^A$ that in turn break a total of $m$ gauge
symmetries \ua.  The equations of motion for the auxiliary field
components $D_a$ of the vector supermultiplets $V_a$ take the form:
\beq D_a = \sum_A K_A q^a_A \phi^A - \delta_{Xa}\ell\dx/2, \quad \dx =
-{\Tr T_X \over 48 \pi^2}, \label{da}\eeq
where $q_A^X$ is the \ux\ charge of the scalar field $\phi^A$: 
$T_X\phi^A = q_A^X\phi^A$ and $\ell$ is the dilaton field introduced
in (\ref{dual}) except that the duality relation in the classical
limit now reads
\beq \ell^{-1} = 2\re s - b{\sum_I}\ln(2\re t^I) +
c_X\dx/2,\label{dual2}\eeq
with $c_X$ the scalar component of the vector superfield $V_X$.  The
derivatives $K_A = \pp K/\pp\phi^A$ of the K\"ahler potential $K =
k(\ell) + G(t + \t,|\phi^A|^2)$ are functions of the real moduli, and
the vacuum conditions $D_a=0$ determine the \vev's of $\phi^A$ as
functions of the dilaton and moduli: $\lvev\phi^A\rvev =
\phi^A(\ell,t+\t)$. The vacuum values $\lvev \ell\rvev$, and $\lvev
t\rvev$ remain undetermined, and SUSY and T-duality remain unbroken at
the scale $\Lambda_D$ where the \ua\, gauge symmetries are broken.
The effective theory obtained by integrating out the massive vector
bosons should reflect these features.  By promoting the conditions
$D_a=0$, with $D_a$ given in (\ref{da}), to superfield equations, it
has been shown\cite{gg} how to construct an effective theory below the
\ux\, breaking scale that has manifest local SUSY and T-duality, and
preserves the correct linearity condition for the linear multiplet
$L$.  This effective theory has several new features: 1) A modified
K\"ahler potential for the dilaton, which can affect dilaton/axion
cosmology, the gaugino/gravitino mass ratio, and the scales of SUSY
breaking and of coupling constant unification. 2) Modified couplings
of moduli to the GS term and to hidden sector matter that govern the
moduli masses. 3) A modified effective K\"ahler metric for matter,
which together with possible \ua\, charges, can affect soft terms in
the scalar potential below the scale of SUSY breaking.  4) Massless
chiral multiplets (``D-moduli'') associated with the large vacuum
degeneracy at the scale $\Lambda_D$ of the D-term induced breaking of
the \ua's that are potentially dangerous for a viable modular
cosmology.\cite{joel}

\subsection{The effective theory below the condensation scale}

The effective theory below the scale of condensation in a strongly
coupled hidden sector with gauge group $\G_c$ was studied\cite{alex}
for a class of models models in which either a minimal set $n=m$ of
scalar fields acquire \vev's $\myvev{\phi^A}\sim \sqrt{\dx}$ that
break the $m$ \ua's, or there are $N$ replicas of minimal sets
$\phi^A$ with identical charges that acquire \vev's.  If in addition
we assume a minimal K\"ahler potential for matter:
\bear K(\Phi,\Ph) = \sum_A x^A + \sum_M x^M, 
 \quad x^{A,M} = e^{G^{A,M} +
2\sum_aq^a_MV_a}|\Phi^{A,M}|^2,\label{minkp}\eear
where by definition $\myvev{\Phi^M}=0$, and the functions $G^{A,M}(t+\t)$ 
assure T-duality of the K\"ahler potential, the masses of the complex
scalars $\phi^M$, that include observable sector particles, 
are given by
\bear m^2_M &=& {m_{3\over2}^2\over1+2z}\[\(1 - \zeta_M{(1+z)^2\over z}\)^2 +
2z - \zeta_M(3+z)\]\nonumber\\
 \zeta_M &=& \sum_{a,A}q^a_M Q^A_a, \quad \sum_a Q^B_a q^a_A =
\delta^B_A, \quad z = b_c\ell.\eear
Note that the D-term contribution to these masses is $(m^2_M)_D = -
\zeta_M z^{-2} m^2_{3\over2}\[1 + O(z)\].$
The leading order ($\sim z^{-2}$) terms linear in $\zeta_M$, ({\it
i.e.} linear in the \ua\, charges $q^a_M$) are canceled by other terms
in the scalar potential. As a result the squared masses are positive,
$m^2_M>0$, over most of relevant parameter space.  If $z\ll 1$ and
$\zeta_M\sim 1$, $m^2_M\gg m^2_{3\over2}$, so the D-terms dominate
over the contribution one gets in their absence ($\zeta^M=0$).  Since
the gaugino masses are unchanged with respect to the model of Section
3, this results in an increased ratio $m_0/m_{1\over2}$ that may be
too large for a viable phenomenology.  There are several possible
cures for this. 1)  If we take instead of (\ref{minkp})
\beq K(\Phi,\Ph) = - \sum_\alpha C_\alpha
\ln\[1 - C_\alpha^{-1}\(\sum_A x^A_\alpha + \sum_M
x^M_\alpha\)\],\label{nonmin}\eeq
There is little change in scalar masses $m^M$, but the effective
K\"ahler metric for $\ell$ is modified in a way that can increase
$m_{1\over2}$; for example, by up to a factor four in the FIQS model
if $C_\alpha = 1$. If we relax the condition $z\ll 1$ we can significantly
reduce the scalar masses. For example in the FIQS models, the smallest
possible squark, slepton and Higgs masses are in the range
$1.5m_{3\over2} \le m_0\le 18m_{3\over2}$ if $\ell = 1$; this reduces
to $.3m_{3\over2} \le m_0\le 5.5m_{3\over2}$ if $\ell = 5$.  Since we
need $b_c\ll1$ to generate a gauge hierarchy, this would suggest
strongish coupling, in other words a point in the Ho\v{r}ava-Witten
scenario that is not quite the WCHS limit. However, in the presence of
string nonperturbative effects\cite{shenk} $\myvev{\ell}$ is not the
coupling constant which is instead given by the \vev\, of a function
$s(\ell)$: $g^{-2} = \myvev{s(\ell)},$ $k'(\ell) - 2\ell s'(\ell),$
where $k(\ell)$ is the dilaton K\"ahler potential, and the second
equality assures a canonical Einstein term.  The model described in
Section 3 requires string nonperturbative effects (SNPE) to stabilize
the potential at strong coupling, that is, to prevent
$V(\ell\to\infty)\to -\infty$.  For example, the parameterization
\bear f(\ell) = 2\ell s(\ell) - 1 = \sum_na_nx^ne^{-x}, \quad x =
\beta/\sqrt{\ell},\label{param}\eear
was used in.\cite{us} A solution was found with vanishing cosmological
constant, weak coupling $g^2\approx.5$ and $f\sim 1$ at the vacuum.
In the presence of D-terms we always have $V(\ell\to\infty)\to +\infty$,
but SNPE are still needed to stabilize the dilaton at weak coupling
and zero cosmological constant:
\beq \myvev{V} \propto \ell k'(\ell) - 3r z/(1 +
z)^2=0,\label{vacen}\eeq
where the parameter $r$ depends on the choice of the K\"ahler
potential for $\phi^A$; $r=1$ for the minimal choice (\ref{minkp}) and
can be larger for a nonminimal choice as in (\ref{nonmin}), {\it e.g.}
$r =$ 1--4 in the FIQS model with $C_\alpha = 1$. However there is now
more freedom in choice of parameterization.  For example, if we take
the dilaton K\"ahler potential
\bear k = - \ln(2s) + \delta k - \ln[1 + h(s)], \quad s = s(\ell),\eear
where the first term gives the classical relation, $\delta k$ is the
contribution from $\myvev{x^A}\sim\dx\ell$, and $h$ is the SNPE
contribution.
If $0< 1 + h = \epsilon \ll 1,$ then $\ell k' \sim
\ell^{-1}\sim\epsilon^{-1}.$
Choices of $h$ similar to the parameterizations (\ref{param}) of $f$ 
used in\cite{us} can give $\ell\approx 5,\quad \ell k' \approx .25$
with $g^2 = s^{-1} = .5$.

The D-moduli couplings to matter condensates lift some of the vacuum
degeneracy at the \ua-breaking scale to give masses to all of the real
scalar D-moduli.  While these are much larger than the gravitino mass
if $z\ll 1$, pushing all of them up to cosmologically safe levels
tends to conflict with the need to reduce the scalar/gaugino mass
ratio in the observable sector in generic models.  In addition one
expects massless D-axions and/or massless D-fermions.  For example in
the FIQS model with three minimal, identically charged sets of of six
fields $\phi^A$ acquiring \vev's to break six \ua's, one linear
combination of these comprises the eaten Goldstone bosons, while the
other two sets of chiral superfields acquire F-term masses such that
the axions remain massless.  In that model there are at least 12
additional states associated with flat directions for which the
complex scalars acquire masses and the fermions do not.  This
particular model is not viable in any case, since it cannot reproduce
the observed SM Yukawa textures,\cite{Giedt:2000} and in
the present context it gives implausibly large values for
$m_{3\over2},\;\Lambda_c$.

Although the D-term modifies the dilaton metric $k'/2\ell$, it
is still  suppressed by the vacuum condition
(\ref{vacen}) if $z\ll1$, giving an
enhanced dilaton mass $m_\sigma$ and a suppressed axion coupling $f_a$.
Because the effective theory above the SUSY-breaking scale is
modular invariant, one again 
obtains moduli stabilized at
self-dual points giving FCNC-free dilaton dominated SUSY-breaking. An
enhancement of the ratio $m_{t^I}/m_{3\over2}$ can result from
couplings to condensates of $U(1)$-charged D-moduli, that also carry
T-modular weights. For example in the FIQS model one gets
$m_\tau^I\approx 10m_{3\over2},$ $m_a^I\approx(2$--$4)m_{3\over2}$.

\section{Conclusions}

	       The message of this talk is three-fold:
1) Quantitative studies with predictions for observable phenomena
are possible within the context of the WCHS.
2) Experiments can place restrictions on the underlying theory,
such as the parameter space of the strongly coupled hidden gauge
sector, as shown in Figure~\ref{fig:abn}, as well as the
superpotential couplings, modular weights and \ua\, charges of
D-moduli when an anomalous $U(1)$ is present.  Experiments can also
inform us about Plank scale physics, such as matter couplings to the
GS term.  The one-loop corrections to the soft scalar potential are
also sensitive to the details of Plank scale physics.
3) Searches for sparticles should avoid restrictive
assumptions, since explicit string-derived models have particle
spectra that do not necessarily conform to conventional scenarios.

\section*{Acknowledgments}
I am indebted to my many collaborators.  This work was supported in
part by the Director, Office of Energy Research, Office of High Energy
and Nuclear Physics, Division of High Energy Physics of the
U.S. Department of Energy under Contract DE-AC03-76SF00098 and in part
by the National Science Foundation under grants PHY-95-14797 and
INT-9910077.


\begin{thebibliography}{99}
\bibitem{refs} For references to the original literature see
M.K. Gaillard, in {\it 2001: A Spacetime Odyssey} (Eds. M.J. Duff and
J.T. Liu, World Scientific 2002) p. 70, hep-ph/0108022.
\bibitem{us1} P. Bin\'{e}truy, M. K. Gaillard and Y.-Y. Wu, {\it
Nucl. Phys.}  B {\bf 481}, 109 (1996); J.A. Casas, \pl B {\bf 384},
103 (1996).
\bibitem{shenk} S.H. Shenker, in {\it Random Surfaces and Quantum
Gravity}, Eds. O. Alvarez, E. Marinari, and P. Windey, NATO
ASI Series B262 (Plenum, NY, 1990); 
T. Banks and M. Dine, {\it Phys. Rev.} D {\bf 50}, 7454
(1994); E. Silverstein, \pl B {\bf 396}, 91 (1997).
\bibitem{us} P. Bin\'{e}truy, M. K. Gaillard and Y.-Y. Wu, {\it
Nucl. Phys.} B {\bf 493}, 27 (1997); {\it
Phys. Lett.} B {\bf 412}, 228 (1997).
\bibitem{gg} M. K. Gaillard, J. Giedt, Nucl. Phys. B 636 (2002) 365
and B 643 (2002) 201.
\bibitem{alex} M.K. Gaillard, J. Giedt and Aleksey Mints,
in preparation.
\bibitem{rs} L. Randall and R. Sundrum, {\it Nucl. Phys.} 
B {\bf 557}, 557 (1999).
\bibitem{hit} G. Giudice, M. Luty, H. Murayama and R. Rattazzi,
{\it JHEP} {\bf 9812}, 027 (1998).
\bibitem{bgn} M.K. Gaillard and B. Nelson, {\it Nucl. Phys.} B {\bf 588},
197 (2000); P. Bin\'etruy, M.K. Gaillard and B. Nelson, {\it Nucl. Phys.}
B {\bf 604}, 32 (2001).
\bibitem{john} J.H. Schwarz, Nucl. Phys. Proc. Suppl. B {\bf 55}, 1
(1997).
\bibitem{mike} M. Green, a seminar.
\bibitem{hw} P. Ho\v rava and E. Witten, \np B {\bf 460},
506 (1996) and B {\bf 475}, 94 (1996).
\bibitem{iban} L.E. Ib\`a\~nez, H.-P. Nilles and F. Quevedo, \pl B
{\bf 187}, 25 (1987); A. Font, L. Ib\`a\~nez, D. Lust and F.  Quevedo,
\pl B {\bf 245}, 401 (1990).
\bibitem{fiqs} A. Font, L. E. Ib\`a\~nez, F. Quevedo and
A. Sierra, {\it Nucl. Phys.} B {\bf 331}, 421 (1990).
\bibitem{nilles}H.P. Nilles, {\it Phys. Lett.} B {\bf 115}, 193 (1982).
\bibitem{thresh} L.J. Dixon, V.S. Kaplunovsky and J. Louis, \np 
B {\bf 355}, 649 (1991); I. Antoniadis, K.S. Narain and T.R. Taylor, \pl 
B {\bf 267}, 37 (1991).
\bibitem{gsterm} G.L. Cardoso and B.A. Ovrut, \np B {\bf 369}, 315
(1993); J.-P. Derendinger, S. Ferrara, C. Kounnas and F. Zwirner, {\it
Nucl. Phys.} B {\bf 372}, 145 (1992).
\bibitem{gnw} M.K. Gaillard, B. Nelson and Y.Y. Wu,
{\it Phys. Lett.} B {\bf 459}, 549 (1999).
\bibitem{gn} M.K. Gaillard and B. Nelson,
{\it Nucl. Phys.} B {\bf 571}, 3 (2000).
\bibitem{abn}  B. Nelson and A. Birkedal-Hansen, {\it Phys. Rev.}
D {\bf 64}, 015008 (2001).
\bibitem{modprob} G.D. Coughlan, W. Fischler, E.W. Kolb, S. Raby and G.G.
 Ross, \pl B {\bf 131}, (1983) 59.
\bibitem{bd} T. Banks and M. Dine, \np B {\bf 505}, 445 (1997).
\bibitem{ad} I. Affleck and M. Dine {\it Nucl. Phys.} B {\bf 249},
361 (1985).
\bibitem{drt} M.Dine, L. Randall, and S. Thomas, {\it Nucl Phys.}
B {\bf 458}, 291 (1996).
\bibitem{gmo} M.K. Gaillard, H. Murayama and K. A. Olive,
{\it Phys. Lett.} B {\bf 355}, 71 (1995).
\bibitem{lyth} M.K. Gaillard, H. Murayama and D.H. Lyth, 
{\it Phys. Rev.} D {\bf 58}, 123505 (1998);
M.K. Gaillard and Mike J. Cai, {\it Phys. Rev.} D {\bf 62}, 047901 (2000).
\bibitem{lsp} M. Kawasaki, T. Moroi, and T. Yanagida, \pl
B {\bf 370}, 52 (1996).
\bibitem{cgmo} M.K. Gaillard, B.A. Campbell, M.K. Gaillard,
H. Murayama and K.A. Olive, {\it Nucl. Phys.} B {\bf 538}, 351 (1999).
\bibitem{efo} J. Ellis, T. Falk, K.A. Olive and M. Schmitt, \pl B {\bf
388}, 97 (1996) and \pl B {\bf 413}, 355 (1997); J. Ellis, T. Falk,
G. Ganis and K.A. Olive {\it Phys. Rev.} D {\bf 62}, 075010 (2000).
\bibitem{joel2} J. Giedt, Ann. of Phys. (N.Y.) 297 (2002) 67.
\bibitem{dsw}  M. Dine, N. Seiberg and E. Witten, \np B {\bf 289},
589 (1987); J. Attick, L. Dixon and A. Sen, \np B {\bf 292}, 109 (1987);
M. Dine, I. Ichinose and N. Seiberg, \np B {\bf 293}, 253 (1988).
\bibitem{joel} F. Buccella, J.P. Derendinger, S. Ferrara and
C.A. Savoy, Phys. Lett. {\bf B115:}375 (1982); M.K. Gaillard and
J. Giedt, {\it Phys. Lett.}  B {\bf 479}, 308 (2000).  
\bibitem{Giedt:2000}
J.~Giedt,
Nucl.\ Phys.\ B {\bf 595} (2001) 3
[Erratum, {\it ibid}. {\bf 632} (2002) 397].
%

\end{thebibliography}
\end{document}